\documentstyle[times,pramana,epsf,floats]{ias}
\def\beq{\begin{equation}}
\def\eeq{\end{equation}}
\def\bea{\begin{eqnarray}}
\def\eea{\end{eqnarray}}
\newcommand{\sla}{\!\!\!\!/ \,}

\newcommand{\vk}{\mbox{\boldmath$k$}}

\begin{document}
\mark{{Van Hove Singularities....}{M. G. Mustafa and M. H. Thoma}}
\title{Can Van Hove singularities be observed in relativistic 
heavy-ion collisions ?}

\author{Munshi G. Mustafa$^1$\thanks{Talk presented} and Markus H. Thoma$^2$}
\address{$^1$ Theory Group, Saha Institute of Nuclear Physics, 
1/AF Bidhan Nagar, Calcutta 700 064, India}
\address{$^2$ Max-Plank-Institut f\"ur Extraterrestrische Physik, 
Giessenbachstra$\beta$e, 85748 Garching, Germany} 
\abstract{
 Based on general arguments the in-medium quark propagator in a
quark-gluon plasma leads to a quark dispersion relation consisting 
of two branches, of which one exhibits a minimum at some finite 
momentum. This results in a vanishing group velocity for collective
quark modes. Important quantities such as the production rate of low 
mass lepton pairs and mesonic correlators depend inversely on this group 
velocity. Therefore these quantities, which follow from self energy diagrams 
containing a quark loop, are strongly affected by Van Hove singularities 
(peaks and gaps). If these sharp structures could be observed in relativistic 
heavy-ion collisions it would reveal the physical picture of the 
QGP as a gas of quasiparticles.}

\maketitle
\section{Introduction}


Recently, Van Hove singularities in the
QGP~\cite{Bra90,Mus99,Mus00,Pes00,Kar01,Tho01}, caused by the collective 
quark modes appearing due to the interaction with the medium, 
have attracted great attention as they could probe the  
QGP as a gas of quasiparticles formed 
in heavy-ion collisions. One of the quark modes, the plasmino branch,
has a minimum at some finite momentum leading to a vanishing group velocity.
Thus the density of states, which is inversely proportional to the group 
velocity, diverges, causing Van Hove singularities 
on relevant quantities like the dilepton production rate and mesonic 
correlators. Here we would like to discuss the role of Van Hove singularities 
in the QGP. 

In the next section, we briefly recall the classic idea of Van Hove 
singularities in the density of states as discussed in the context of
condensed matter physics. We will demonstrate in the following two sections
that Van Hove singularities appear as a dynamical aspect of QGP on account 
of the effective quark propagator containing the quark self-energy.
In sec.3 we will demonstrate this through the hadronic correlators using 
the hard thermal loop (HTL) resummed qurak propagator whereas in sec. 4  
through the nonperturbative dilepton production rate, using an effective
quark propagator obtained by taking into account the nonperturbative 
gluon condensate above the critical temperature derived from lattice QCD. 
In sec. 5
we will argue that the minimum in the plasmino branch and thus the
appearance of Van Hove singularities is a general property of massless 
fermions at finite temperature. Finally, in sec. 6 we will discuss
the possibilities and difficulties to observe Van Hove singularities which 
could reflect the physical picture of the
QGP as a gas of quasiparticles produced in heavy-ion collisions.

\section{Van Hove Singularities}

In condensed matter physics Van Hove singularities in the density of states
were discussed on structural aspects of a solid 
by Van Hove~\cite{van} in 1953. The density of state of a system ({\it e.g}. 
phonons, electrons) is given as
\begin{equation}
 g(\omega)=\sum_n \int \frac{{\rm d}^3k}{(2\pi )^3}\> \delta
(\omega -\omega_n(\vk)), \label{eq1}
\end{equation}
implying the number of allowed wave vectors ${\vk}$ 
in the level $n$ having energy $\omega_n(\vk)$
within the energy interval $\omega$ and $\omega +{\rm d}\omega$. By surface
integration it can be given~\cite{ash} as
\beq
g(\omega)=\sum_n \int \frac{{\rm d}S_\omega}{(2\pi )^3}\>
\frac{1}{|{\bf \nabla} \omega_n(\vk)|}, \label{eq2}
\eeq
which is explicitly related to the group velocity, 
${|{\bf \nabla} \omega_n(\vk)|}$, which is the 
derivative of the dispersion relation. 
Due to symmetries in a crystal
the group velocity vanishes at certain momenta, resulting in a divergent
integrand in (\ref{eq2}). This divergence is integrable in 3-dimensions,
leading to a finite density of states. In lower dimensions, however,
Van Hove singularities appear. For example, a 2-dimensional electron
gas shows logarithmic singularities, which have been discussed in connection
with high-$T_c$ superconductors \cite{Mar97}. A 2-dimensional electron gas
has also been discussed widely in the context of the quantum Hall effect. 
As discussed here, the Van Hove singularities are linked to the structural
aspect of a solid whereas, as we will see below, in the QGP they are 
due to dynamical reasons. 

\section{Thermal Hadronic Correlaton Function}

\subsection{Definition}

Meson correlators are constructed from meson currents
$J_M (\tau,\vec{x}) =\bar{q}(\tau, \vec{x})\Gamma_M q(\tau, \vec{x})$,
where $\Gamma_M = 1$, $\gamma_5$, $\gamma_\mu$, $\gamma_\mu \gamma_5$ 
for scalar, pseudo-scalar, vector and pseudo-vector channels, respectively.
The thermal two-point functions in coordinate space, $G_M(\tau,\vec{x})$,
are defined as
\begin{eqnarray}
G_M(\tau,\vec{x}) &=&
\langle J_M (\tau, \vec{x}) J_M^{\dagger} (0, \vec{0}) \rangle
\nonumber \\
&=& T \sum_{n=-\infty}^{\infty} \int
{{\rm d}^3p \over (2 \pi)^3} \;{\rm e}^{-i(\omega_n \tau- \vec{p} \vec{x})}\;
\chi_M(\omega_n,\vec{p})~~,
\label{eq3}
\end{eqnarray}
where $\tau \in [0,1/T]$, and the Fourier transformed correlation function
$\chi_M(\omega_n,\vec{p})$ is given at the discrete Matsubara modes,
$\omega_n = 2n \pi T$. The imaginary part of the momentum space correlator
gives the spectral function $\sigma_M(\omega,\vec{p})$,
\begin{equation}
\chi_M(\omega_n,\vec{p}) = -\int_{-\infty}^{\infty} {\rm d}
\omega {\sigma_M(\omega,\vec{p}) \over i\omega_n - \omega +i\epsilon}
\Rightarrow 
\sigma_M(\omega,\vec{p}) = {1\over \pi} {\rm Im}\;  \chi_M(\omega,\vec{p}).
\label{eq4}
\end{equation}
Using eqs.~(\ref{eq3}) and (\ref{eq4}) we obtain
the spectral representation of the thermal correlation function in
coordinate space at fixed momentum ($\beta =1/T$),
\begin{equation}
G_M(\tau,\vec{p}) =  \int_{0}^{\infty} {\rm d} \omega\;
\sigma_M (\omega,\vec{p})\;
{{\rm cosh}(\omega (\tau - \beta/2)) \over {\rm sinh} (\omega \beta/2)}~~.
\label{eq5}
\end{equation}

\subsection{Hard Thermal Loop (HTL) Approximation}
Perturbative QCD at finite temperature is based on the fact that the
temperature dependent running coupling constant is small at high
temperatures due to asymptotic freedom, $T\rightarrow \infty
\Rightarrow \alpha _s(T)=g^2/4\pi \rightarrow 0$. At a typical temperature
of $T=250$ MeV we expect $\alpha _s= 0.3$ -- 0.5. This suggests that
perturbative theory could work at least qualitatively.

However, restricting only to bare propagators and vertices perturbative 
QCD can lead to serious problems, {\it i.e.}, infrared divergent and gauge 
dependent results 
for physical quantities, {\it e.g.}, in the case of
the damping rate of a long wave, 
collective gluon mode in the QGP. The sign and magnitude of the gluon
damping rate was found to be strongly gauge dependent~\cite{Kaj85,Lop85}.
The reason for such behaviour is the fact that bare perturbative QCD at 
finite temperature is incomplete, {\it i.e.}, higher order diagrams missing
in bare perturbation theory can contribute to lower order in coupling constant.
In order to overcome these problems, the HTL resummation technique, 
{\it an improved 
perturbation theory}, has been suggested by Braaten and 
Pisarski~\cite{Bra90a} and also by Frenkel and 
Taylor~\cite{Fre90}, in which those diagrams can be taken
into account by resummation. The following prescription has been suggested  
for the improved perturbation theory:

\begin{enumerate}
\item Isolate those diagrams which should be resummed. The starting point is
the separation of scales in the weak coupling limit, $g<<1$, since there are
two momentum scales in a plasma
of massless particles: {\bf (i)} {\it hard}, where the 
momentum $\sim$ temperature, $T$ and {\bf (ii)} {\it soft}, where the momentum 
$\sim $ thermal 
mass ($\sim \ gT$, $g<<1$). These diagrams are HTL self energies and 
vertices 
given by one loop diagram, where the external momenta are soft and the 
loop momenta are hard. For example, in the case of self energies in QED or QCD
a result proportional to $g^2T^2$ was obtained, which is equivalent to the 
high temperature approximation, $T>> p, \ \ |p_0|$, found earlier by 
Klimov ~\cite{Kli82} and Weldon~\cite{Wel82}, and also to the semiclassical
approximation by Silin~\cite{Sil60}.   
\item Effective propagators in the HTL approximation are constructed using the 
Dyson-Schwinger equations. The HTL vertex follows simply by adding the HTL 
correction to the bare vertex. This effective vertex 
is also related, except for 
scalar particles, to the HTL propagator through Ward identity.
\item These effective quantities, propagators and vertices, 
can then be used as in ordinary perturbation theory if {\it all} 
legs are soft, leading to results that are complete to 
leading order in the coupling constant, gauge independent and improved in
the infrared behaviour. If, on the other hand,  at least one leg of the 
propagtors or vertices is hard, bare quantities are sufficient. 
\end{enumerate}

\subsection{Quark self energy in HTL-approximation}
\vspace{-0.3in}
\begin{figure}[htbp]
\epsfxsize=8cm
\hspace*{-3cm}\centerline{\epsfbox{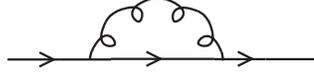}}
\caption{Quark self energy having quark momentum $K$.}
\label{fig:qself}
\end{figure}
The most general ansatz for the fermionic self energy 
in rest frame of the plasma is given by~\cite{Wel82a}  

\beq
\Sigma(K)=-a(k_0,k)K\sla -b(k_0,k)\gamma_0 \ , \label{eq6}
\eeq
where $K=(k_0,{\bf k})$,  $k=|{\bf k}|$ and the quark mass is neglected
assuming that the temperature is much larger than the quark mass, which holds 
at least for $u$ and $d$ quarks.  The scalar quantities $a$ and $b$ are
given by the traces over self energies in figure~\ref{fig:qself} as, 
respectively,
\bea
a(k_0,k) & = & \frac{1}{4k^2}\> \left [{\rm {tr}}\, (K\sla \Sigma ) - k_0\>
{\rm{tr}}\, (\gamma _0
\Sigma )\right ],\nonumber \\
b(k_0,k) & = & \frac{1}{4k^2}\> \left [K^2\> {\rm{tr}}\, (\gamma _0 \Sigma )
 - k_0\> {\rm{tr}}\,
(K\sla \Sigma )\right ],\label{eq7}
\eea
and obtained in HTL-approximation as
\bea
a(k_0,k)&=&\frac{m_q^2}{k^2}\> \left (1-\frac{k_0}{2k}\> \ln
\frac{k_0+k}{k_0-k}\right )\ \ , \nonumber \\
b(k_0,k)&=&\frac{m_q^2}{k^2}\> \left (-k_0+\frac{k_0^2-k^2}{2k}\> \ln
\frac{k_0+k}{k_0-k}\right )\ \ ,\label{eq8}
\eea
where $m_q^2={g^2T^2}/{6}$, is the effective quark mass. The quark self 
energy in eq.(\ref{eq6}) has an imaginary part below the light cone,
$k_0^2-k^2<0$, representing Landau damping for space like quark momenta.
Furthermore, the general ansatz in eq.(\ref{eq6}) is also chirally 
invariant in spite of the appearance of an effective quark mass.

\subsection{Effective Propagators and Vertices in HTL-approximation}
\begin{figure}[htbp]
\epsfxsize=8cm
\centerline{\epsfbox{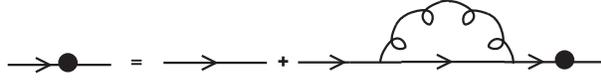}}
\caption{Effective quark propagator in HTL-approximation}
\label{fig:qprop}
\end{figure}
Resumming the quark self-energy by using
the Dyson-Schwinger equation, the effective quark propagator
in figure~\ref{fig:qprop} can be written as 
\beq
S(K)=[K\sla -\Sigma(K)]^{-1} . \label{eq9}
\eeq
For massless quarks, decomposing the effective quark propagator in 
(\ref{eq9}) into helicity eigenstates, we get
\beq
S(K)=\frac{\gamma_0-\hat {\bf k}\cdot {\vec \gamma}}{2D_+(K)}+
\frac{\gamma_0+\hat {\bf k}\cdot {\vec \gamma}}{2D_-(K)} \ \ , \label{eq10}
\eeq
with
\beq
D_\pm(k_0,k)=(1+a(k_0,k))\> (-k_0\pm k)+b(k_0,k) \ \ \ . \label{eq11}
\eeq
The relevant QED like HTL-vertex is related to this propagator
through the Ward identity~\cite{Bra90,Fre90} 
\beq
K_\mu \Gamma^\mu(K_1,K_2;K)=S^{-1}(K_1)-S^{-1}(K_2) \ \ . \label{eq14}
\eeq
\vspace{-0.3in}
\begin{figure}[htbp]
\epsfxsize=6cm
\centerline{\epsfbox{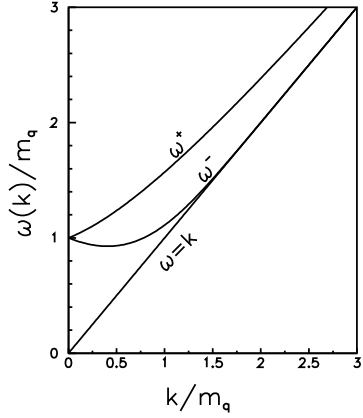}}
\vspace{-0.4in}
\caption{Quark dispersion relation in HTL-approximation along with the free 
($\omega=k$) one.}
\label{fig:hdisp}
\end{figure}

Now, the zeros of $D_\pm(K)$ give the in-medium dispersion relation 
for quarks. As shown in figure~\ref{fig:hdisp} the upper curve 
$\omega_{+}(k)$ corresponds to the solution of $D_{+}(K)=0$, whereas 
the lower curve $\omega_{-}(k)$ represents the solution of $D_{-}(K)=0$. 
Both branches start from a common effective
mass, $m_q$, obtained in the $k\rightarrow 0$ limit~\cite{Bra90}.
The $\omega_{+}(k)$ branch describes the propagation of an ordinary quark
with thermal mass, and
the ratio of its chirality to helicity is $+1$. On the other hand,
the $\omega_{-}(k)$ branch corresponds to the propagation of a quark mode
with a negative chirality to helicity ratio. This branch represents the
plasmino mode which is absent in the vacuum but appears as consequence of 
the medium due to the broken Lorentz invariance, and has a shallow 
minimum.  This corresponds to a purely collective long 
wave-length mode, whose spectral strength decreases exponentially at high 
momenta.  
For high momenta, however, both 
branches approach the free dispersion relation. 

\subsection{Hadronic spectral and correlation function in HTL-approximation}

The hadronic spectral functions, $\sigma_M(\omega, p)$ in (\ref{eq4}),
of the temporal correlators are proportional to the imaginary part of 
the quark loop diagram. Here we will calculate the imaginary part of
the quark loop diagram corresponding to the vector meson channel,
which is related to the static dilepton production rate
(${\vec p}=0$) as~\cite{Bra90} 
\beq
\sigma_V(\omega)=\frac{1}{\pi}{\rm{Im}}\ \chi_V(\omega)
=\frac{18\pi^2N_C}{5\alpha^2}
\left( e^{\beta \omega}-1\right ) \omega^2 \frac{dR}{d^4xd\omega d^3p} 
({\vec p}=0) . \label{eq15} 
\eeq
For the details of the other channels, {\it e.g.}, pseudovector, scalar 
and pseudoscalar, see Refs.~\cite{Kar01,Tho01}. 
\begin{figure}[htbp]
\epsfxsize=4cm
\centerline{\epsfbox{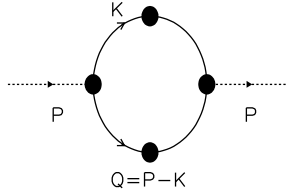}}
\caption{Self-energy diagram involving quarks loop in HTL-approximation.}
\label{fig:hself}
\end{figure}
Now the vector meson correlation function can be written 
from figure~\ref{fig:hself} as
\begin{equation}
\hspace{-0.7in}
\chi_V \left(\omega, p\right) =2N_{C}T\sum _{k_{0}}\int \frac{d^{3}k}
{(2\pi )^{3}}{\rm{tr}}\left[ S\left( K\right) \Gamma ^{\mu}(-Q,-K;P)
S \left( Q\right) \Gamma ^{\nu}(Q,K;-P) \right] \, ,
\label{eq16}
\end{equation}
where, \( Q=P-K \), $N_C=3$ and $\chi_V$ 
denotes the trace over the Lorentz indices of $\chi_M^{\mu\nu}$. Using
the effective quantities from (\ref{eq10}) and (\ref{eq14}), 
eq.~(\ref{eq15}) at ${\vec p}=0$ can be written as
\bea
\sigma_V(\omega )&=&\frac{4N_c}{{\pi^2}}(e^{\beta\omega}-1)\int ^{\infty }_{0}
dk\, k^{2}\int _{-\infty }^{+\infty }dx \int ^{+\infty }_{-\infty }
dx ^{\prime } n_{F}\left( x \right) n_{F}\left( x^{\prime }\right)
\delta \left( \omega-x -x^{\prime }\right) \nonumber \\
&& \hspace{-0.5in}
\times \left\{ 4\left( 1-\frac{x^{2}-x^{\prime 2}}
{2k\omega}\right) ^{2} \rho _{+}\left(x ,k\right) \rho _{-}
\left(x^{\prime },k\right) 
 + \left( 1+\frac{x^{2}+x^{\prime 2}-2k^{2}-2m_{q}^{2}}
{2k\omega} \right) ^{2}\right.\nonumber \\
&&\hspace{-0.5in}
\left. \times \rho _{+}\left(x,k\right) \rho _{+}
\left(x^{\prime },k\right)
 +\left( 1-\frac{x^{2}+x^{\prime 2}-2k^{2}-2m_{q}^{2}}
{2k\omega}\right) ^{2}\rho _{-}\left(x,k\right) \rho _{-}(x^\prime, k) 
\right. \nonumber\\
&& \hspace{-0.6in} \left. 
 + \theta \left( k^{2}-x^{2}\right) \frac{m^{2}_{q}}
{4k\omega^{2}} \left( 1-\frac{x^{2}}{k^{2}}\right)
 \left[ \left( 1+\frac{x}{k}\right)
\rho _{+}\left(x^{\prime },k\right) +\left( 1-\frac{x}{k}\right)
\rho _{-}\left(x^{\prime },k\right) \right] \  \right\}, \label{eq17}
\eea
where $x$ and $x^\prime$ are the energies of the internal particles, $n_F$
is the fermi distribution function, and $\rho_\pm$ are the spectral function
for collective quark modes in the medium~\cite{Bra90} having a {\it pole} 
contribution from the in-medium dispersion relation and a {\it cut} 
contribution for space like quark momenta related to Landau damping below 
the light cone ($k_0^2 <k^2$). 
The meson spectral function, constructed from two quark propagators,
will have pole-pole, pole-cut and 
cut-cut contributions,
$\sigma_M(\omega)=\sigma^{\rm{pp}}(\omega)+\sigma^{\rm{pc}}(\omega)
+\sigma^{\rm{cc}}(\omega)$. 
Here we quote only the pole-pole expression for vector mesons,
\bea
\sigma_V^{\rm{pp}} & = & \frac{1}{4m^{4}_{q}}\left[ {k^{2}_{1}}n^{2}_{F}
\left( \omega _{+}\left( k_{1}\right) \right) \frac{\left( \omega ^{2}_{+}
\left( k_{1}\right) -k^{2}_{1}\right) ^{2}}{2\omega_+^\prime(k_1)}
\left( 1+\frac{\omega ^{2}_{+}\left( k_{1}\right) -k^{2}_{1}-m^{2}_{q}}
{k_{1}\omega }\right) ^{2}\right. \nonumber \\
 & + & \left. k^{2}_{2}n\, ^{2}_{F}\left( \omega _{-}\left( k_{2}\right) 
\right) \left( 1-\frac{\omega ^{2}_{-}\left( k_{2}\right) -k^{2}_{2}
-m^{2}_{q}}{k_{2}\omega }\right) ^{2}
\frac{\left( \omega ^{2}_{-}\left( k_{2}\right) -k^{2}_{2}\right) ^{2}}
{2\omega_-^\prime(k_2)}\right. \nonumber \\
&+&\left.  2k^{2}_{3}\, n_{F}\left( \omega _{+}\left( k_{3}\right) \right) 
n_{F}\left( -\omega _{-} \left( k_{3}\right) \right) 
\frac{\left( \omega ^{2}_{+} \left( k_{3}\right) -k^{2}_{3}\right) 
\left( \omega ^{2}_{-}\left( k_{3}\right) -k^{2}_{3}\right)} 
{\omega_+^\prime(k_3)-\omega_-^\prime(k_3)}
\right. \nonumber \\
 & \times  & \left. 
\left( 1+\frac{\omega ^{2}_{+}\left( k_{3}\right) +\omega ^{2}_{-}
\left( k_{3}\right) -2k^{2}_{3}-2m^{2}_{q}}{2k_{3}\omega }\right) ^{2}
\right] \ , \label{eq19}
\eea
where $k_1$ is the solution of $\omega-2\omega_+(k)=0$, and $k_2^i$ and
$k_3^i$ are the solutions of $\omega-2\omega_-(k)=0$ and 
$\omega-\omega_+(k)+\omega_-(k)=0$, respectively, which can have two 
solutions at low momenta due to the plasmino mode. Each term in (\ref{eq19})
corresponds to different physical processes~\cite{Bra90,Kar01,Tho01} and is
inversely proportional to the derivative of the dispersion relation 
(group velocity). 
The other contributions ($\sigma^{\rm{pc}}$ and $\sigma^{\rm{cc}}$) 
can easily be obtained~\cite{Bra90,Kar01,Tho01} from (\ref{eq17}). 
Similar results for the scalar channel have also been derived 
and detailed discussions are given in Ref.~\cite{Kar01,Tho01}. 
Free meson spectral functions are also 
obtained using bare propagators and vertices
in figure~\ref{fig:hself} as~\cite{Kar01,Flo94}
\beq
\sigma ^{{\rm free}}_{M}\left( \omega \right)
=\frac{N_{C}}{4\pi ^{2}} \omega ^{2}\tanh \left( \frac{\omega }{4T}\right)
a_{M}\ \ , \label{eq20}
\eeq
where $a_M= (+)-1$ for (pseudo)scalar and $(-)+2$ for (pseudo)vector.
\begin{figure}[htbp]
\centerline{{\epsfxsize=6cm {\epsfbox{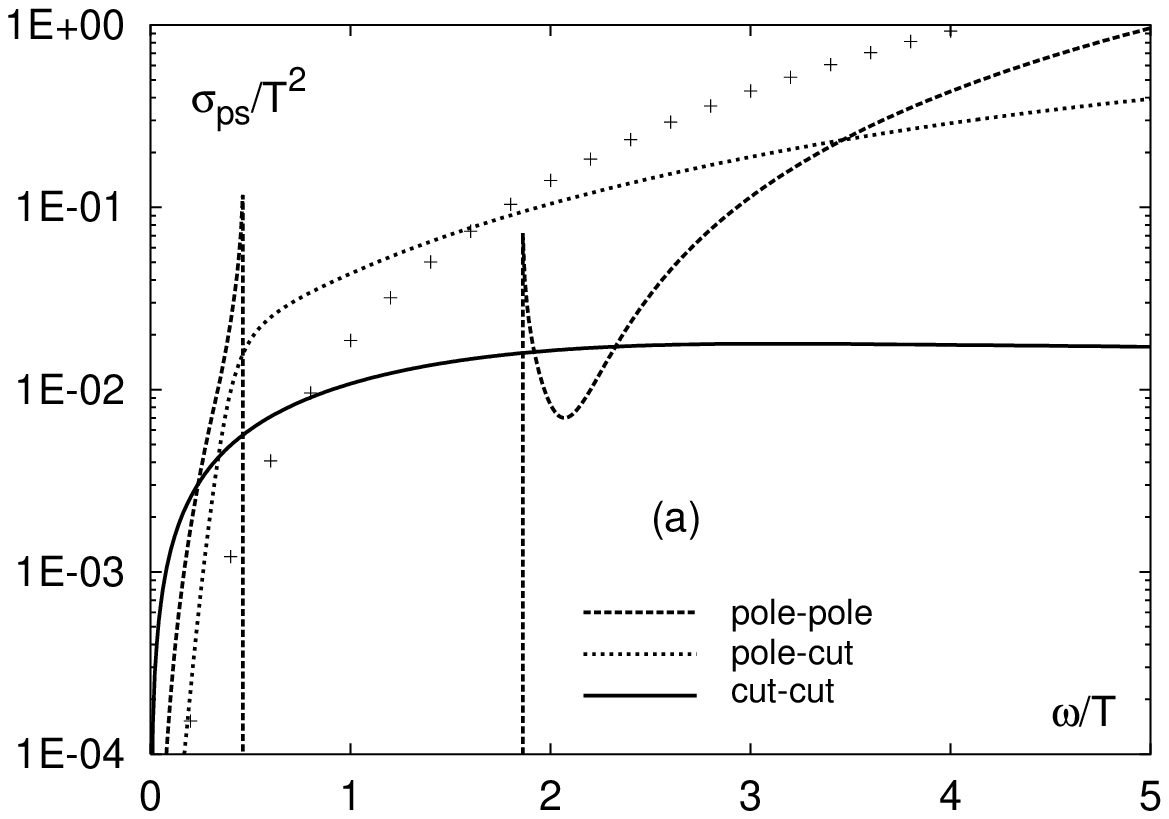}}}
{\epsfxsize=6cm {\epsfbox{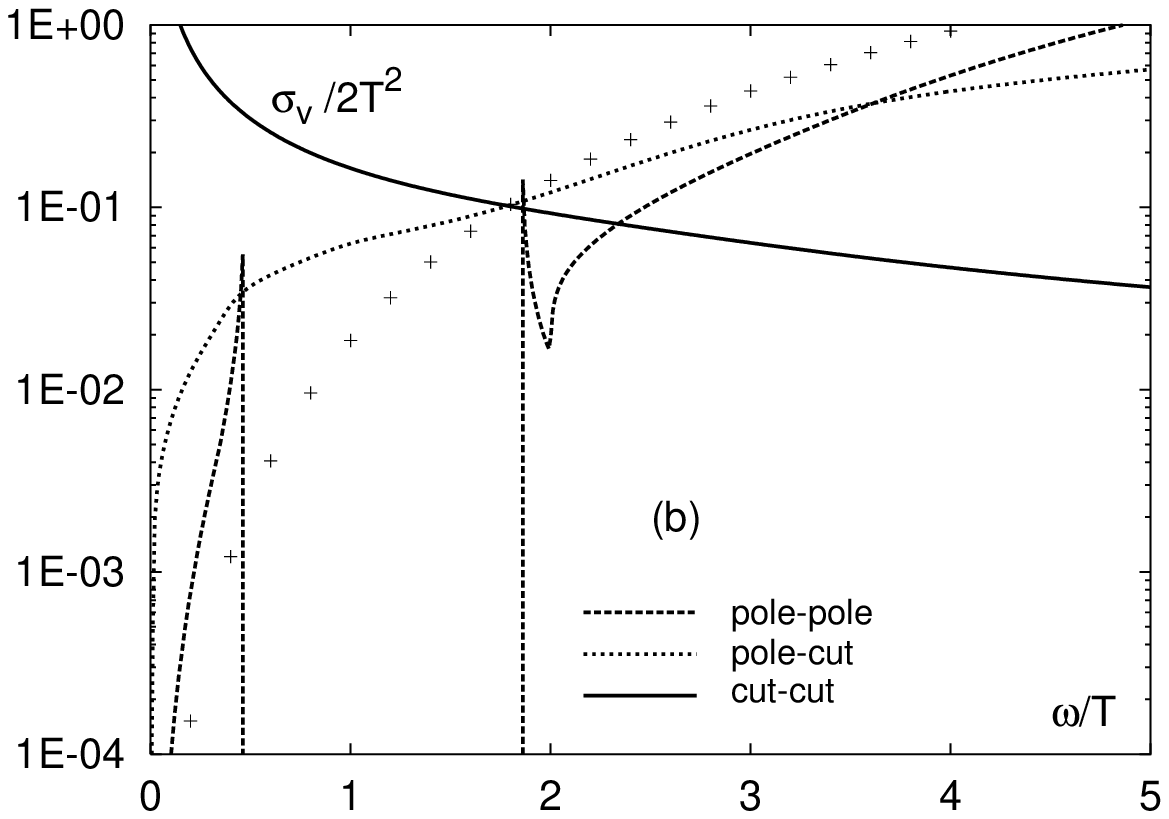}}}}
\caption{
(a) Pseudoscalar and (b) vector meson spectral functions in HTL approximation
for $m_q/T=1$. The free one is represented by crosses.}
\label{fig:hspec}
\end{figure}

In figure~\ref{fig:hspec} we show the different contributions for 
pseudoscalar and vector mesons for $m_q/T=1$, {\it i.e.}, $g=\sqrt 6$.
Generically the pole-pole term of the meson spectral function in 
HTL-approximation describes three physical processes: {\bf (1)} the 
annihilation of collective quarks ({\it e.g.}, 1$^{\rm{st}}$ term 
in~(\ref{eq19})), 
{\bf (2)} the annihilation of two plasminos (2$^{\rm{nd}}$ 
term in~(\ref{eq19}))
and {\bf (3)} the transition from upper to lower branch (3$^{\rm {rd}}$ term
in~(\ref{eq19})).
The transition process starts at
zero energy and continues until the maximum difference $\omega =0.47\> m_q$
between the two branches at $k_3=1.18\> m_q$. At this point a Van Hove
singularity is encountered due to the vanishing denominator
$\omega'_+(k_3)-\omega'_-(k_3) =0$, implying a diverging density of states.
The plasmino annihilation starts at $\omega = 1.86\> m_q$ with another Van 
Hove singularity corresponding to the minimum of the plasmino branch at 
$k_3=0.41\> m_q$, where $\omega'_-(k_2) = 0$. This contribution falls off 
rapidly due to the exponentially suppressed spectral strength of the plasmino 
mode for large energies, where only the first process, quark-antiquark 
annihilation starting at $\omega =2m_q$, contributes. For large energies this
dominates and approaches the free results (crosses) for $\omega >> m_q$.
The pole-cut and cut-cut contributions, which involve 
external gluons as can be seen by cutting the HTL quark self energy, lead
to a smooth contribution to the spectral function. The 
pole-pole and pole-cut contributions in both channels are of similar
magnitude,
while the cut-cut contribution at small $\omega$ vanishes in the pseudoscalar 
channel but diverges in the vector channel. This has its origin in the 
structure of HTL quark-meson vertex~\cite{Bra90,Fre90}, which is 
required for the 
vector channel containing a collinear singularity~\cite{Bai94}, 
whereas bare vertices are sufficient for the (pseudo)scalar channel.
This implies that
higher order diagrams in the HTL expansion contribute to the same order 
in the coupling constant~\cite{Aur98}, which, of course, indicates that the 
low frequency part of the vector meson spectral function is inherently 
nonperturbative. As discussed, the vector meson spectral function is
linked to the dilepton production~\cite{Bra90,Kar01} at high temperature 
whereas the pseudoscalar correlator is related to 
the chiral condensate and thus to the chiral 
susceptibility~\cite{Hat94,Cha02}. 
\begin{figure}[htbp]
\centerline{{\epsfxsize=6cm {\epsfbox{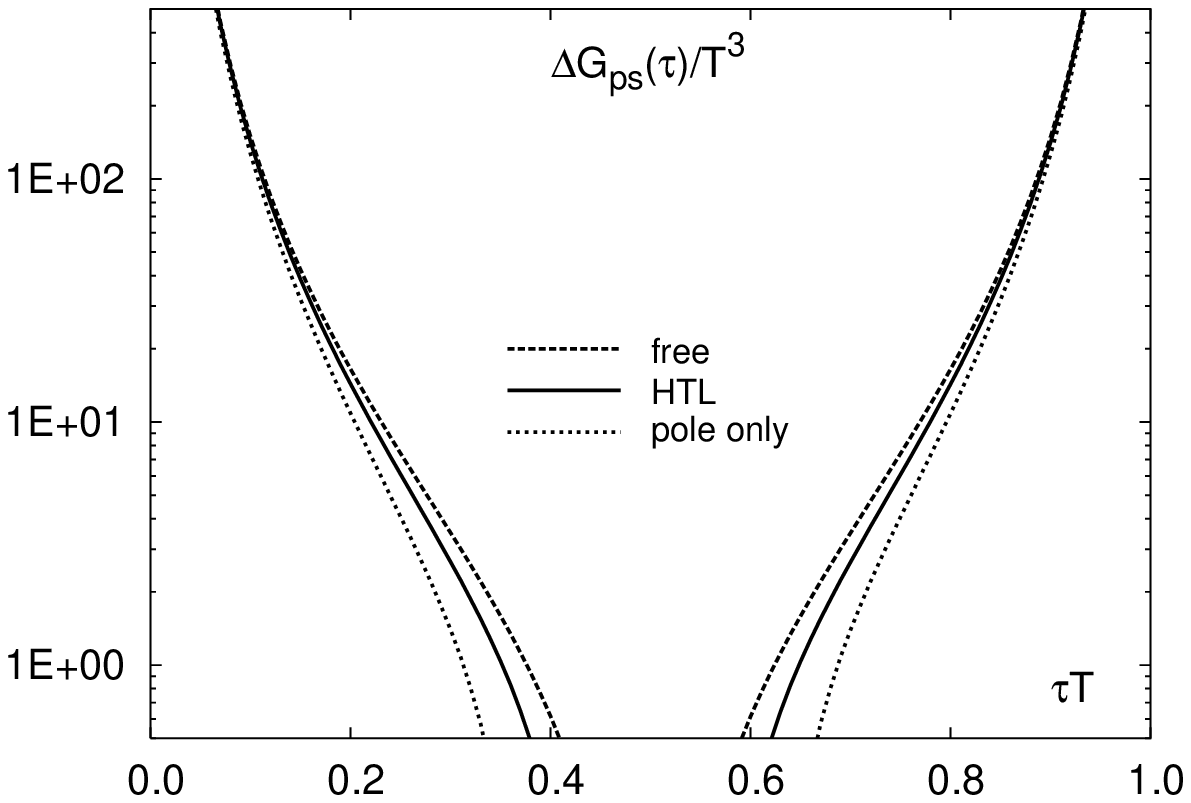}}}
{\epsfxsize=6cm {\epsfbox{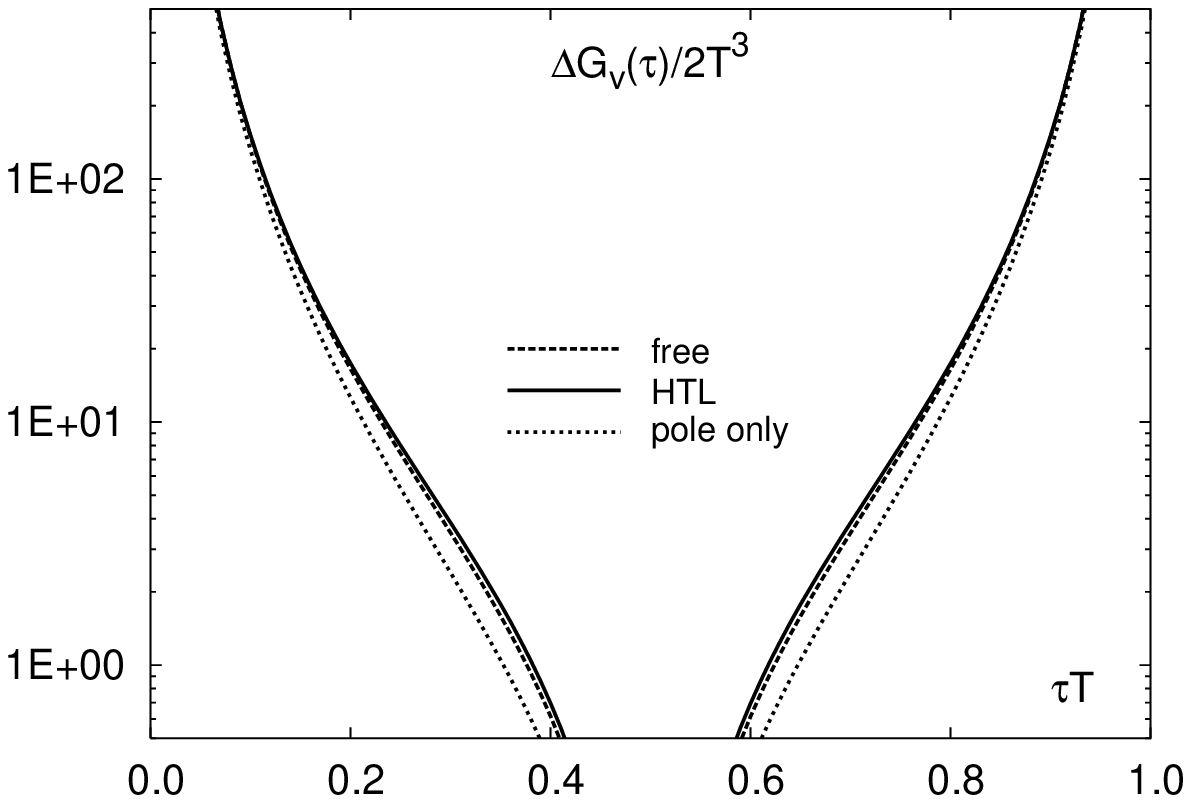}}}}
\caption{ The subtracted thermal pseudoscalar (left) and  vector (right) 
meson correlation functions in HTL-approximation for $m_q/T=2$.}
\label{fig:hdel}
\end{figure}

The temporal correlators for pseudoscalar and vector mesons
can be obtained from their spectral function according to (\ref{eq5}). 
The competing effects of pole and cut contributions 
in the spectral function carries over to the correlation functions,  
which is found to be almost similar~\cite{Kar01,Tho01}, unlike in
lattice calculations~\cite{Lae02},
to the free one with bare propagators and vertices. 
A linear divergence, due to the effective HTL-vertices, of the spectral 
function in the vector channel at low frequencies also 
renders the temporal correlator infrared divergent. 
Although the pseudoscalar correlation functions are infrared
finite, the low frequency part of the spectral functions 
will also be modified significantly from contributions of higher order 
diagrams. It is reasonable to consider modified correlation functions,
$\Delta {G}_M (\tau) \equiv {G}_M (\tau) - {G}_M (\beta/2)$,
which are less sensitive to details of the low frequency part of the
spectral functions. In the subtracted correlation functions the infrared 
divergences are eliminated. As shown in figure~\ref{fig:hdel} they are 
well-defined in both channels.
One observes that after elimination of the infrared divergent 
parts the structure of the pole and cut contributions is similar in the 
scalar and vector channel. The vector correlator seems to be even closer 
to the leading order perturbative (free) correlator than the scalar 
one.

\section{Gluon condensate and nonperturbative dilepton production}

It seems that quite generally nonperturbative effects~\cite{Boy96} 
dominate the temperature regime attainable in heavy-ion collisions
and the application of perturbative calculations there becomes questionable
as the coupling constant is not small. Such nonperturbative effects were 
made explicit by lattice QCD calculations~\cite{Lae02}
below and above the phase transition temperature. The gluon condensate, which
appears due to the broken scale invariance at zero as well as at finite 
temperature, describes the nonperturbative imprints of the QCD ground 
states and have extensively been used for studying hadron properties
at zero and finite temperature. Lattice QCD so far is not capable to
compute dynamical quantities. As an alternative method to lattice 
and perturbative QCD it has been suggested to include the gluon condensate 
obtained from lattice QCD into the parton propagators at zero~\cite{Lav88} 
as well at finite~\cite{Tho99} temperature.
In this way, nonperturbative effects are taken into account within
the Green functions technique, which then can be exploited to calculate
various physical quantities like the dilepton rate, hadronic correlators,
susceptibilities, etc.

\subsection{Quark propagation }

\begin{figure}[htbp]
\epsfxsize=4cm
\centerline{\epsfbox{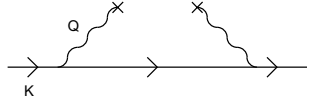}}
\caption{Quark self-energy diagram in the presence of the gluon condensate.}
\label{fig:gcself}
\end{figure}
\begin{figure}[htbp]
\centerline{{\epsfxsize=5cm {\epsfbox{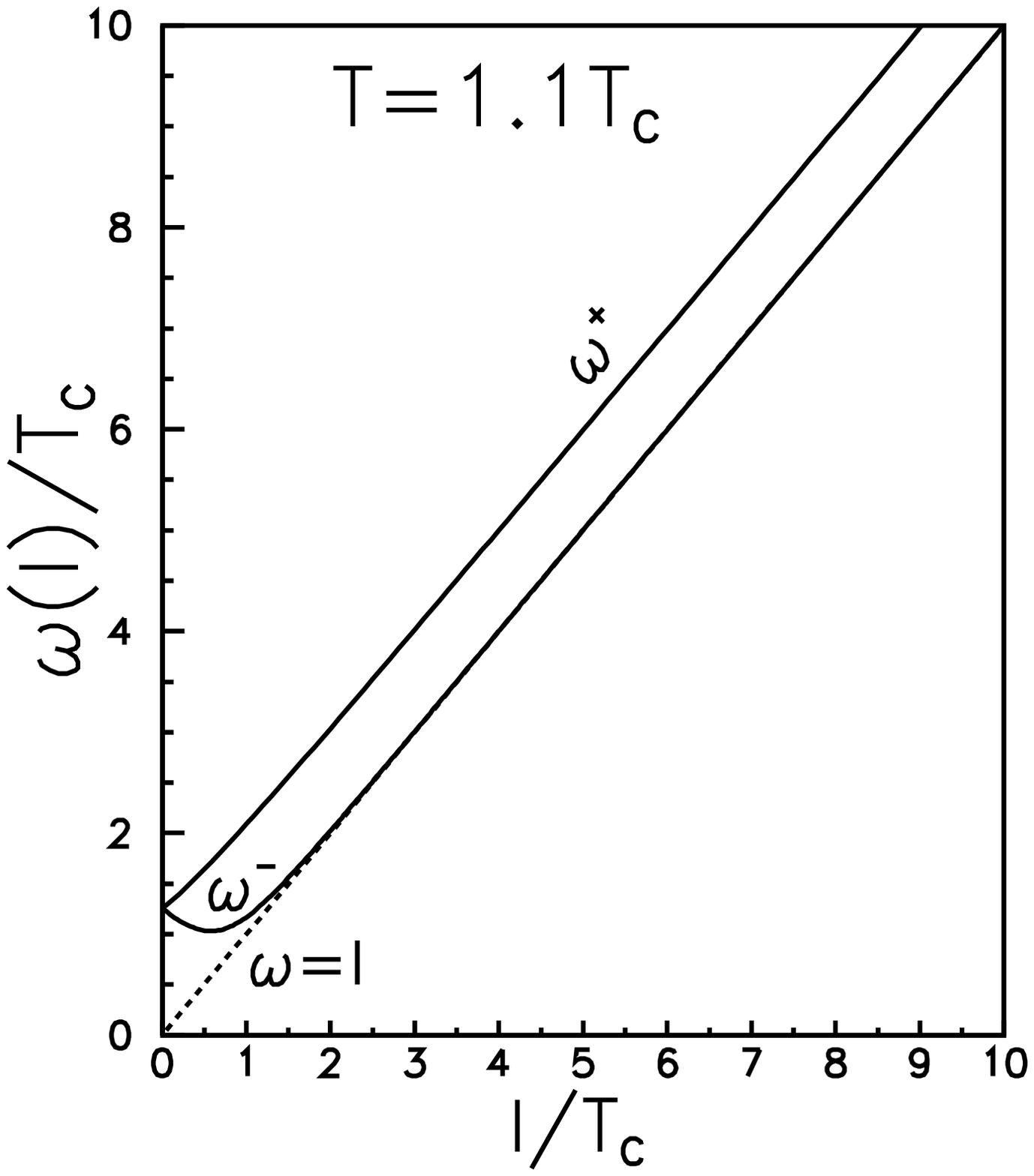}}}
{\epsfxsize=5cm {\epsfbox{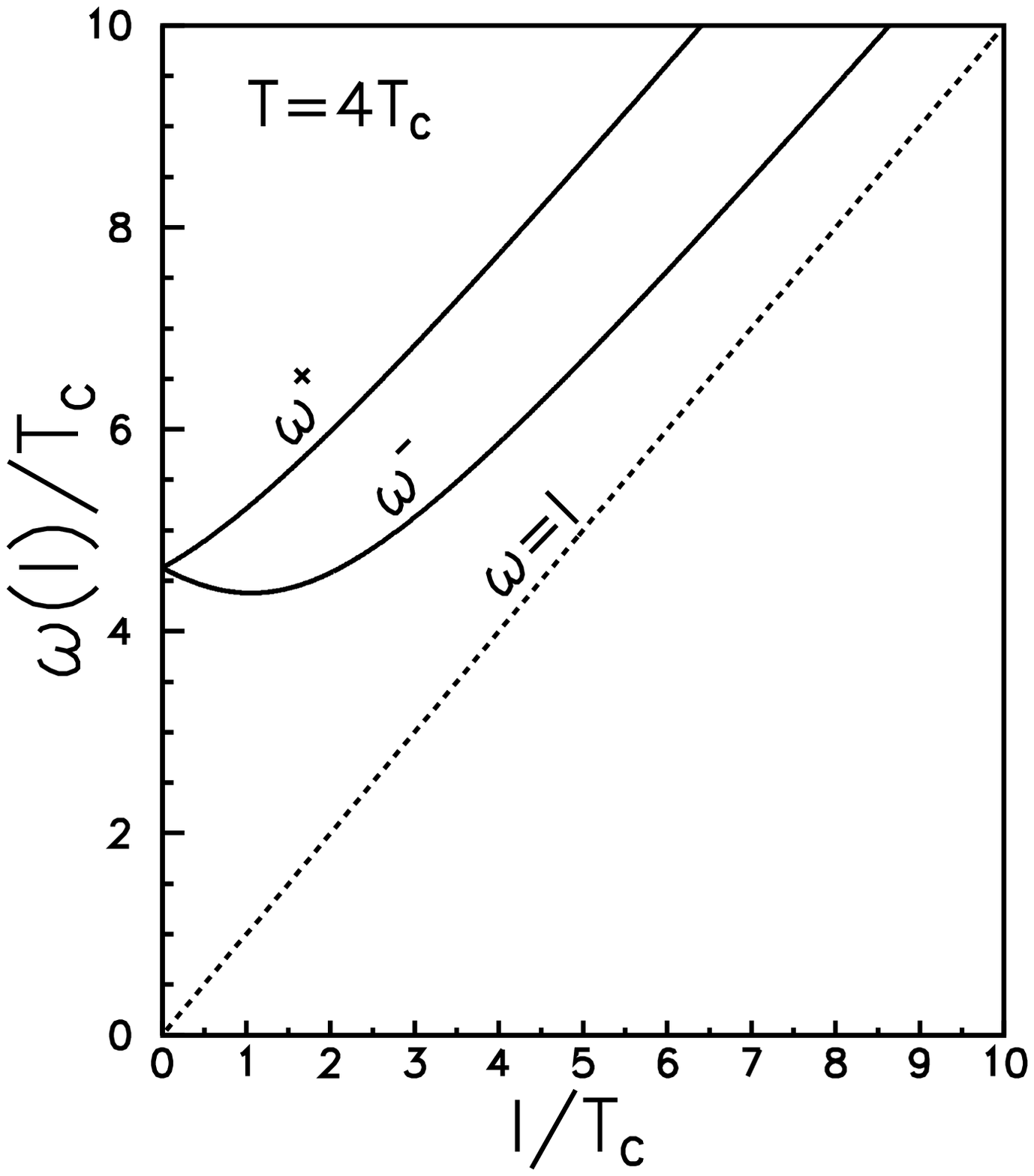}}}}
\vspace{-0.4in}
\caption{Quark dispersion relation in the presence of the gluon condensate with
$T=1.1T_c$ (left) and $T=4T_c$ (right). Here, read l=$k$. }
\label{fig:gdisp}
\end{figure}

Now, the functions $a(k_0,k)$ and $b(k_0,k)$ in the general 
effective quark propagator at finite temperature defined in (\ref{eq10}) 
and (\ref{eq11}) can be obtained using self energy diagram with the 
gluon condensate in figure~\ref{fig:gcself} as~\cite{Mus99,Mus00,Tho99}
\begin{eqnarray}
a &=& - \frac{g^2}{6} \frac {1}{K^6} \biggl [ \left ( \frac {1}{3} k^2
- \frac{5}{3}k_0^2\right )
\langle {\mathcal E}^2 \rangle_T
 - \left ( \frac{1}{5}k^2
- k_0^2 \right ) \langle {\mathcal B}^2 \rangle_T \biggr ] , \nonumber \\
b &=& - \frac{4}{9}g^2 \frac {k_0}{K^6} \left [ k_0^2\langle {\mathcal E}^2
\rangle_T
+ \frac{1}{5}k^2\langle {\mathcal B}^2 \rangle_T \right ] ,
\label{eq21}
\end{eqnarray}
with the chromoelectric and and chromomagnetic condensates 
expressed~\cite{Mus99,Mus00,Tho99} 
in terms of the space like ($\Delta_\sigma$) and time like ($\Delta_\tau$)
plaquette expectation values as 
\bea
\frac{\alpha_s}{\pi} \langle {\cal E}^2\rangle_T &=& \frac{4}{11} T^4
\Delta_\tau
- \frac{2}{11} \langle G^2\rangle_{T=0} \ , \nonumber \\
\frac{\alpha_s}{\pi} \langle {\cal B}^2\rangle_T &=& -\frac{4}{11} T^4
\Delta_\sigma
+ \frac{2}{11} \langle G^2\rangle_{T=0} \ , \label{plaq}
\eea
where $\langle G^2\rangle_{T=0}=(2.5\pm 1.0)T_c^4$ and the plaqutte values 
are taken from lattice calculations~\cite{Boy96}. 
This propagator is also related to an
effective vertex~\cite{Mus00a} through the 
QED like Ward identity. The dispersion 
relation in the presence of the 
gluon condensate is shown in figure~\ref{fig:gdisp}, 
which shows the same qualitative features as in the HTL case 
(figure~\ref{fig:hdisp}) 
although it is a completely different approximation. 
However, the thermal mass of the quark is $m_{\rm{eff}}\sim 1.15T$, which is 
independent 
of the coupling constant unlike in the 
HTL case. 

\subsection{Nonperturbative dilepton production}

\begin{figure}[htbp]
\vspace{-0.3in}
\centerline{{\epsfxsize=5cm {\epsfbox{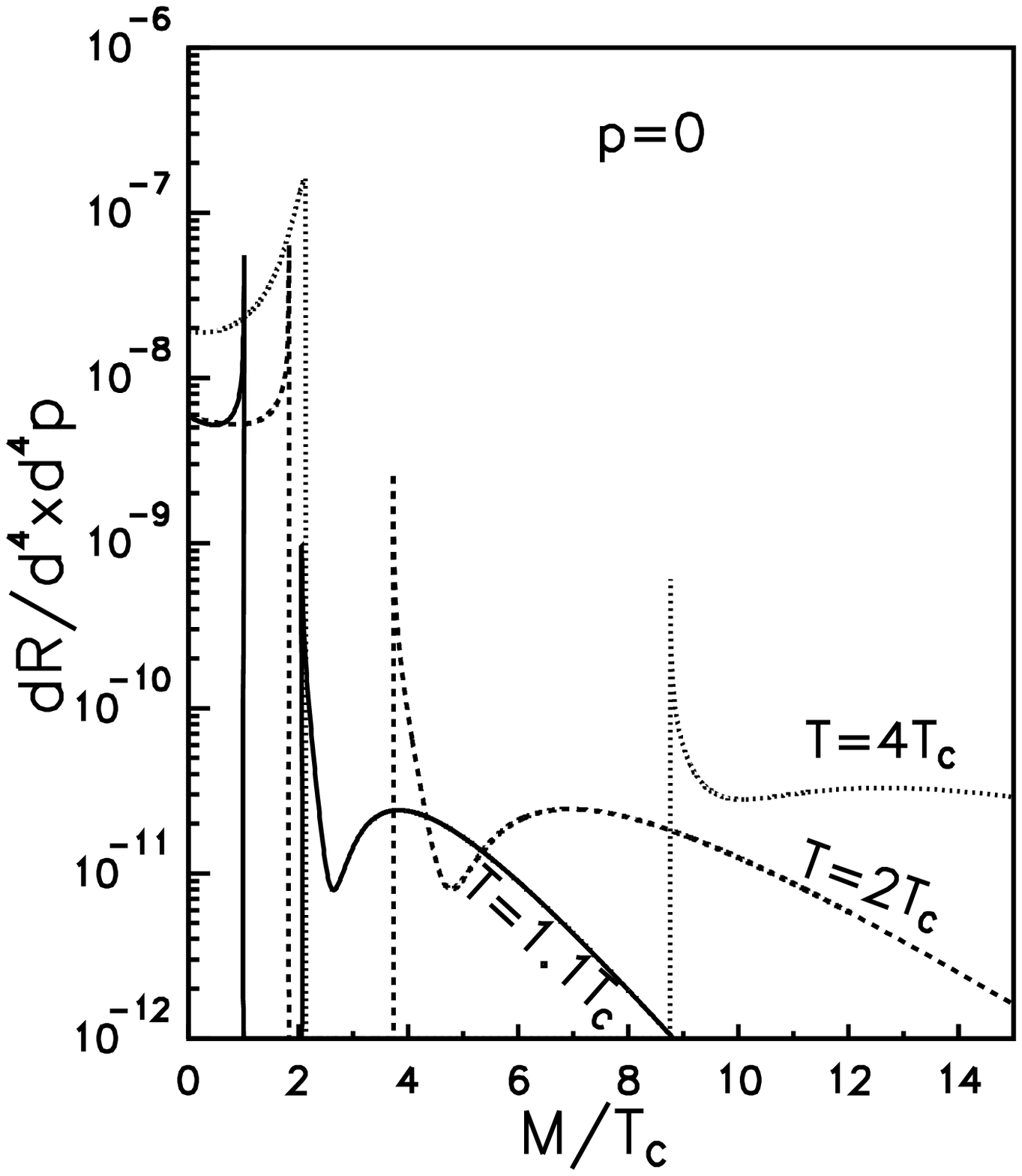}}}
{\epsfxsize=5cm {\epsfbox{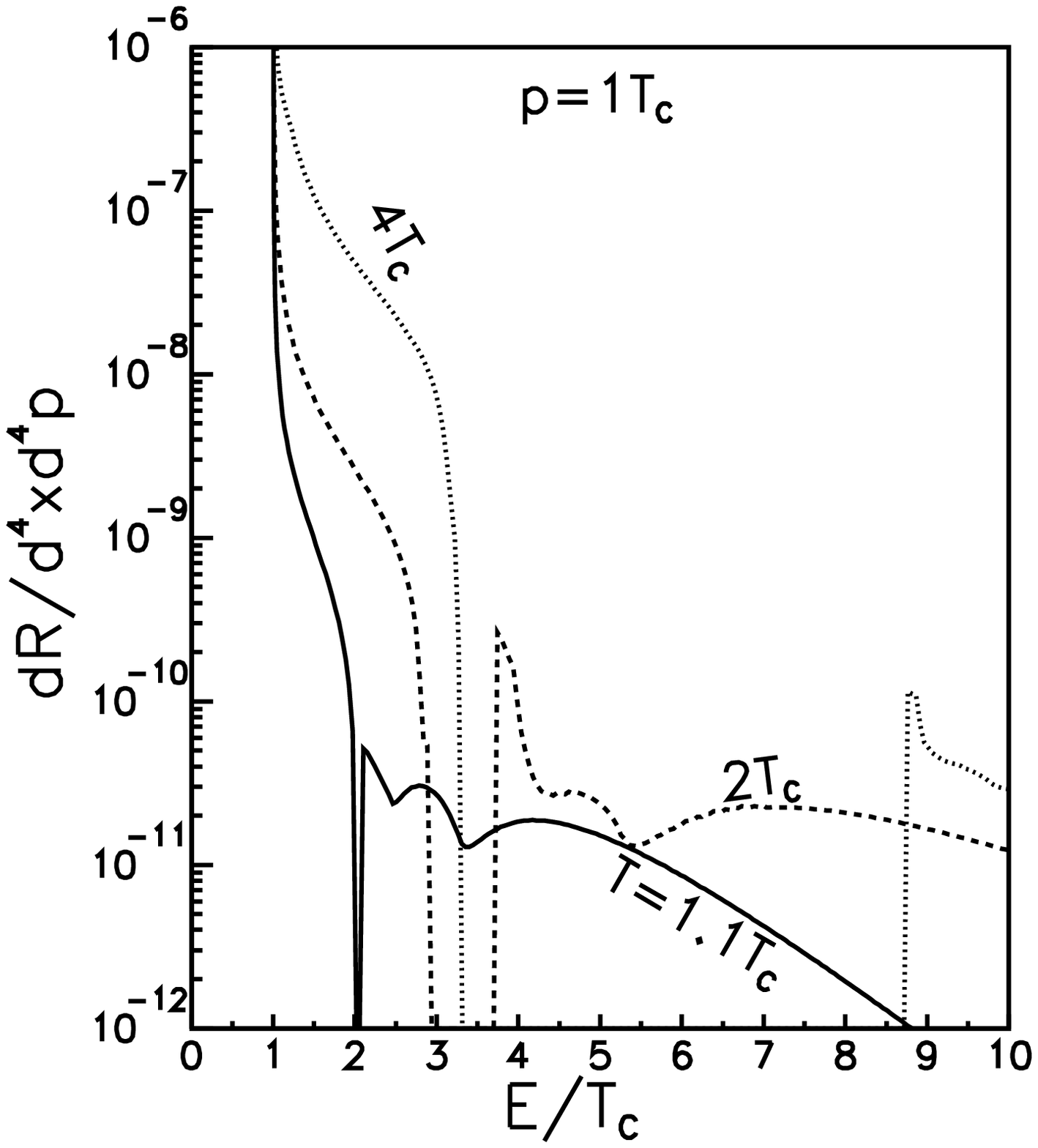}}}}
\vspace{-0.4in}
\caption{Dilepton production in the presence of the gluon condensate
 for $T=1.1T_c$, $2T_c$ and $4T_c$.}
\label{fig:gdist}
\end{figure}
The dilepton production rate using the gluon condensate can be obtained from 
the imaginary part of the photon self energy related to (\ref{eq15}) 
\begin{equation}
\frac{dR}{d^4xd\omega d^3p}=\frac{1}{6\pi^4}\> \frac{\alpha }{M^2}\frac{1}
{e^{\omega/T}-1}\> {\rm{Im}}\Pi_\mu^\mu (P),
\label{eq22}
\end{equation}
where $E=\omega=\sqrt{p^2+M^2}$ is the energy of the virtual photon with 
invariant mass $M$ 
and momentum $p$, and $\alpha$ is the fine structure constant.
The nonperturbative dilepton production rate has been 
obtained in 
Ref.~\cite{Mus99,Mus00} neglecting an 
effective quark-photon vertex containing
the gluon condensate. The results are shown in figure~\ref{fig:gdist} for 
$p=0$ (left) and $p=T_c$ (right) for different temperatures. As in the case 
of soft dileptons~\cite{Bra90} and hadronic correlators~\cite{Kar01,Tho01} 
calculated within the HTL approximation, we also find peaks and gaps in 
the dilepton rate. The peaks (Van Hove singularities) are caused
again by the presence of the minimum in the plasmino dispersion 
(figure~\ref{fig:gdisp}). 
The contribution at small $M$ ending with a Van Hove
singularity comes from an electromagnetic transition from the upper branch
to the lower branch ($q_+ \rightarrow \gamma^* q_-$) of the quark dispersion 
relation. After this peak
there is a gap before the channel for plasmino annihilation
($q_-\> \bar q_-\rightarrow \gamma^*$) opens up
with another singularity. This contribution drops quickly but at 
$M=2m_{\rm{eff}}$
the contribution from the annihilation of collective quarks
($q_+\> \bar q_+\rightarrow \gamma^*$) sets in.
This contribution dominates and approaches the one coming from the
annihilation of bare quarks (Born term \cite{Cle87}) at large $M$.
The peaks and gaps are smoothed out to
some extent for nonzero $p$ as shown in the right panel.
It should be noted that there are no smooth cut contributions
in contrast to the HTL dilepton rate, because the
quark self energy containing the gluon condensate (figure~\ref{fig:gcself}) 
has no imaginary part.

\section{General quark dispersion relation}

It has been demonstrated in the preceding two sections 
with two completely different approximations that the in-medium 
dispersion relation reveals identical qualitative features:
{\bf (1)} there are always two branches;
{\bf (2)} both of them lie above the free dispersion curve ($\omega =k$) and 
start at the same energy, {\it i.e.}, effective mass, in the isotropic 
limit ($k\rightarrow 0$);
{\bf (3)} the two branches have opposite slopes in the isotropic limit; 
{\bf (4)} as shown in the two independent examples, both branches
approach the free dispersion relation thus the plasmino branch 
must always have a minimum.  
Considering the most general ansatz for the fermion self energy and thus for
the quark propagator, 
it has been argued~\cite{Pes00} that the plasmino minimum is a general feature
responsible for the appearance of the van Hove singularities in the
dilepton production rate and hadronic correlators.

\section{Discussions and conclusion}

Considering two independent approximations, the HTL and the 
gluon condensate, for the
quark propagator, we have studied the in-medium quark dispersion relation.
The dispersion relation, in both approximations, is found to govern general
characteristics implying the physical picture of the QGP as a gas of 
quasiparticles. The important feature, which appears to be  
general, is the minimum in plasmino branch due to the broken Lorentz 
invariance. 

One loop dispersive curves are gauge invariant and
whether they are physically observable is, again, a highly nontrivial question.
Future lattice calculation might be able to investigate the 
dispersion relation~\cite{Kar}.
However, it is easy to measure explicitly photon dispersive curves in a QED 
plasma by studying the propagation of classical electromagnetic waves, 
which measures the electric charge density as a function of time and spatial 
coordinates and thus determine the frequency and wave vectors.

But due to confinement there is no such thing in nature like color
fields and also device which could measure the  color charge density. 
What one can measure are correlators of colorless currents like mesonic 
correlation functions and the dilepton production rate. As seen, they will be
affected by the in-medium dispersion relation because a colorless current is
always coupled to a pair of colored particles through their Green's functions
which are  related to the characteristics of the dispersion relation only in a 
indirect way. Also, the thermal modification of vertices is important as 
they are related to the thermal Green's functions through the Ward identity. 

We have calculated the dilepton rates and hadronic correlators using
the in-medium dispersion relation obtained in two different approximation  
and have found in both cases Van Hove singularities, which are due to the 
general characteristic (plasmino mode) of the dispersion relation. 
However, to some extent these structures will be smeared out by 
consideration of finite momentum, higher order processes and the space-time 
evolution of the fireball. If these
structures are observed in the dilepton spectrum or in hadronic
correlation functions in relativistic heavy-ion collisions, 
they would certainly reveal the general nature
of the dispersion relation implying the physical picture of the QGP as
a gas of quasiparticles. Such a signal is not expected from the
hadronic phase~\cite{Rapp00}.
These effects are interesting topics of future investigations. 

\acknowledgments
Most of the works were done in collaboration with
F. Karsch, A. Peshier and A. Sch\"afer.



\begin{thebibliography}{99}
\bibitem{Bra90} E. Braaten, R.D. Pisarski, and T.C. Yuan, Phys. Rev. Lett.
{\bf 64}, (1990) 2242.
\bibitem{Mus99} M. G. Mustafa, A Sch\"afer, and M. H. Thoma, Nucl. Phys. A
{\bf 661} (1999) 653c.
\bibitem{Mus00} M. G. Mustafa, A. Sch\"afer and M. H. Thoma, Phys. Rev.
C {\bf 61} (2000) 024902.
\bibitem{Pes00} A. Peshier and M.H. Thoma, Phys. Rev. Lett. {\bf 84}
 (2000) 841.
\bibitem{Kar01} F. Karsch, M. Mustafa, and M.H. Thoma, Phys. Lett. B {\bf 497}
(2001) 249.
\bibitem{Tho01} M. H. Thoma, Nucl. Phys. (Proc. Suppl.) B {\bf 92} (2001) 162. 
\bibitem{van} L. Van Hove, Phys. Rev. {\bf 89} (1953) 1189.
\bibitem{ash} N.W. Ashcroft and N.D. Mermin, {\sf Solid State Physics}
(Saunders College, Philadelphia, 1976).
\bibitem{Mar97} R.S. Markiewicz, J. Phys. Chem. Sol. {\bf 58} (1997) 1179.
\bibitem{Kaj85} K. Kajantie and J. I. Kapusta, Ann. Phys. (N.Y.) {\bf 160}
(1985) 477.
\bibitem{Lop85} J. C. Parikh, P. J. Siemens, and J.A. Lopez, 
Pramana {\bf 32} (1989) 555.
\bibitem{Bra90a} E. Braaten and R.D. Pisarski, Nucl. Phys. B {\bf 337}
 (1990) 569.
\bibitem{Fre90} J. Frenkel and J. C. Taylor, Nucl. Phys. B {\bf 334}
(1990) 199.
\bibitem{Kli82} V.V. Klimov, Sov. Phys. JETP {\bf 55} (1982) 199.
\bibitem{Wel82}  H.A. Weldon, Phys. Rev. D {\bf 26} (1982) 1394.
\bibitem{Sil60} V.P Silin, Sov. Phys. JETP {\bf 11} (1960) 1136.
\bibitem{Wel82a} H.A. Weldon, Phys. Rev. D {\bf 26} (1982) 2789.
\bibitem{Flo94}W. Florkowski and B.L. Friman, Z. Phys. A {\bf 347}
 (1994) 271.
\bibitem{Bai94} R. Baier, S. Peign\'{e}, and D. Schiff, 
Z. Phys. C {\bf 62} (1994) 337.
\bibitem{Aur98} P. Aurenche {\it et al.}, Phys. Rev. D {\bf 61}
(2000) 116001.
\bibitem{Hat94} T. Hatsuda and T. Kunihiro, Phys. Rep. {\bf 247} (1994) 221.
\bibitem{Cha02} P. Chakraborty, M. G. Mustafa and M. H. Thoma, under 
preparation.
\bibitem{Lae02} E. Laermann, in this procedings.
\bibitem{Boy96} G. Boyd et al., Nucl. Phys. B {\bf 469}, 419 (1996).
\bibitem{Lav88} M. J. Lavelle and M. Schaden, Phys. Lett. 
B {\bf 208} (1988) 419.
\bibitem{Tho99} A. Sch{\"a}fer and M. H. Thoma, Phys. Lett. 
B {\bf 451} (1999) 195.
\bibitem{Mus00a} M. G. Mustafa, A. Sch\"afer and M. H. Thoma, Phys. Lett.
B {\bf 472} (2000) 402 .
\bibitem{Cle87} J. Cleymans, J. Fingberg, and K. Redlich, Phys. Rev. 
D {\bf  35} (1987) 2153.
\bibitem{Kar} F. Karsch, private communication.
\bibitem{Rapp00} R. Rapp and J. Wambach, Adv. Nucl. Phys. {\bf 25} (2000) 1.
\end{thebibliography}
\end{document}